\documentclass[aps,pra,reprint,amsmath,amssymb,showpacs,floatfix]{revtex4-1}

\usepackage[colorlinks,allcolors=blue,hyperindex,breaklinks]{hyperref}

\usepackage{graphicx,units}

\newcommand{\ket}[1]{\ensuremath{\vert{#1\rangle}}} 
\newcommand{\bra}[1]{\ensuremath{{\langle #1}\vert}}

\newcommand{\ketbra}[2]{\ensuremath{|{#1 \rangle}{\langle #2}|}}
\newcommand{\op}[1]{\hat{#1}}
\newcommand{\D}{\text{d}}
\newcommand{\I}{\text{i}}
\newcommand{\E}{\text{e}}
\providecommand{\abs}[1]{\left\lvert#1\right\rvert}

\begin{document}

\title{Performance advantage of protective quantum measurements}

\author{Maximilian Schlosshauer}

\affiliation{Department of Physics, University of Portland, 5000 North Willamette Boulevard, Portland, Oregon 97203, USA}

\begin{abstract}
 We compare the performance of protective quantum measurements to that of standard projective measurements. Performance is quantified in terms of the uncertainty in the measured expectation value. We derive an expression for the relative performance of these two types of quantum measurements and show explicitly that protective measurements can provide a significant performance advantage over standard projective measurements. \\[.2cm]
Journal reference: \emph{Phys.\ Rev.\ A} {\bf 110}, 032215 (2024),  DOI: \href{https://doi.org/10.1103/PhysRevA.110.032215}{\texttt{10.1103/PhysRevA.110.032215}}
\end{abstract}

\maketitle

\section{Introduction}

It is a fundamental tenet of quantum mechanics that measurements will, in general, change the state of the measured system. This may be rigorously quantified in terms of a tradeoff between information gain and disturbance \cite{Fuchs:1996:jj}. Protective quantum measurements \cite{Aharonov:1993:qa,Aharonov:1993:jm,Anandan:1993:uu,Dass:1999:az,Vaidman:2009:po,Gao:2014:cu,Genovese:2017:zz,Piacentini:2017:oo,Rebufello:2021:km,Chen:2023:jn} are noteworthy in this context because they can provide information about expectation values while leaving the state of the measured quantum system (approximately or exactly) unchanged. This is accomplished by combining a weak system--apparatus coupling with an appropriate state-protection procedure. The state protection is based either on the adiabatic theorem \cite{Aharonov:1993:qa} or on the quantum Zeno effect \cite{Misra:1977:aa,Itano:1990:lm,Aharonov:1993:jm,Home:1997:za,Virzi:2022:aa}. In the latter approach, which we shall refer to as a Zeno protective measurement from here on, the system is subjected to a sequence of repeated steps consisting of a weak system--apparatus interaction followed by a projection onto the initial (premeasurement) state. In what follows, we will refer to each such step as a Zeno stage. Experimental realizations of Zeno protective measurements using photons, with up to nine Zeno stages, have been reported in Refs.~\cite{Piacentini:2017:oo,Rebufello:2021:km,Chen:2023:jn}. 

In a Zeno protective measurement, the pointer wave packet shifts deterministically by an amount proportional to the expectation value. This means that the expectation value can be directly read off from the pointer shift, rather than having to be reconstructed as a statistical average over many projective (strong) measurements on an ensemble. While the relevance of protective measurements to foundational issues such as the question of the ontological status of the quantum state is a subject of ongoing debate \cite{Unruh:1994:ll,Rovelli:1994:ll,Ghose:1995:km,Ariano:1996:om,Dass:1999:az,Uffink:1999:zz,Gao:2013:st,Schlosshauer:2014:tp,Gao:2015:km,Combes:2018:lm,Gao:2022:zz}, protective measurements may offer practical advantages over other types of quantum measurements (such as projective measurements on an ensemble) for determining expectation values. Indeed, the existence of a performance advantage of this kind was described in Ref.~\cite{Piacentini:2017:oo}, where the uncertainty in the protectively measured expectation value was reported to be typically smaller than the uncertainty in the expectation value obtained from projective measurements, given comparable resources in both measurement schemes. Using the same performance quantifier (albeit with a modification to buffer the effect of large fiber-optic losses), the advantage was also experimentally observed in Ref.~\cite{Chen:2023:jn}. While Ref.~\cite{Piacentini:2017:oo} gave some partial mathematical results, the performance advantage was reported only in graphical form, and no complete analytical calculations or closed-form results were provided.

In the present paper, we close this gap. We adapt the exact solution of the system--apparatus evolution for a Zeno protective measurement given in Ref.~\cite{Combes:2018:lm} (where it was presented with the different aim of challenging claims regarding the ability of protective measurements to suggest the reality of the quantum state) to derive analytical expressions for the performance advantage provided by protective measurements. We focus on Zeno protective measurements in this paper, as it is the type of protective measurement that has been studied most extensively and is also the only version of a protective measurement for which experimental realizations have been reported \cite{Piacentini:2017:oo,Rebufello:2021:km,Chen:2023:jn}.

This paper is organized as follows. In Sec.~\ref{sec:model}, we define the measurement model and present the relevant results for the quantum-state evolution for this model. In Sec.~\ref{sec:comp}, we use these results to derive an expression for the performance quantifier and analyze the resulting performance advantage of protective measurements. We discuss our findings in Sec.~\ref{sec:discussion}.

\section{\label{sec:model}Model and state evolution}

\subsection{Zeno measurement model}

We consider a two-outcome observable $\op{O} = \op{\Pi}_+ - \op{\Pi}_-$ with eigenstates $\ket{\pm}$. An experimentally relevant example would be the linear polarization observable $\op{O} = \ketbra{H}{H} - \ketbra{V}{V}$ used in the protective-measurement experiments reported in Refs.~\cite{Piacentini:2017:oo,Rebufello:2021:km,Chen:2023:jn}. Let the interaction Hamiltonian be $\op{H}_\text{int} = g \op{O} \otimes \op{P}$, where $g$ is the coupling strength and $\op{P}$ is the operator that generates the translation of the apparatus pointer in the conjugate variable $\op{Q}$ with eigenstates $\{\ket{Q}\}$. The pointer variable $\op{Q}$ may correspond to spatial position, but it may also represent other quantities, such as the photon arrival time used in the experiment in Ref.~\cite{Chen:2023:jn}. We take the total measurement time $T$ (consisting of all Zeno stages) to be fixed and divide it into $N$ Zeno stages of duration $\Delta t$, so that $T=N\Delta t$. It is customary to relate the measurement strength $g$ to $T$ via $g = 1/T$ in order to obtain a pointer shift that is normalized to the interval $[-1,1]$ and that is independent of the number of stages. In this case, the extremal pointer-shift values $-1$ and $+1$ correspond to the measured system being in the eigenstates $\ket{-}$ and $\ket{+}$ of $\op{\Pi}_-$ and $\op{\Pi}_+$, respectively.

The measurement is weak when the shift per Zeno stage is small compared with the width $\sigma$ of the apparatus pointer wave packet in the variable $\op{Q}$. Since the shift per stage is on the order of $1/N$, the weak-measurement regime corresponds to $1/N \ll \sigma$. Clearly, enlarging $\sigma$ or $N$ will both make the measurement weaker. This also provides a method for a reasonable choice of $\sigma$ for a given number of stages.

Let $\ket{\psi}=\sqrt{r}\ket{+}+\sqrt{1-r} \ket{-}$ be the initial state of the system that is to be protected. (Given that the relative phase between the two state components is not accessible in a measurement of the observable $\op{O} = \op{\Pi}_+ - \op{\Pi}_-$, we can disregard it here.) Then each Zeno stage consists of a unitary system--apparatus coupling $\op{U}$ acting over a time $\Delta t$, 
\begin{equation}\label{eq:bvhb111}
  \op{U} = \exp\left( - \I \op{H}_\text{int} \Delta t \right)  = \exp\left[ -  \frac{\I}{N} (\op{\Pi}_+ - \op{\Pi}_-) \otimes \op{P} \right] ,
\end{equation}
followed by a projection $\op{\Pi}_\psi = \ketbra{\psi}{\psi}$ onto
the initial state. Let $\ket{\Phi}$ be the initial state of the apparatus pointer. Then the final combined system--pointer state at the end of $N$ Zeno stages, using the $\op{Q}$ representation for the pointer state, is
\begin{equation}
  \label{eq:hdsv}
  \bra{Q}  \left(\op{\Pi}_\psi \op{U}\right)^N \ket{\psi} \ket{\Phi} = \ket{\psi} \bra{Q}\left(\bra{\psi} \op{U} \ket{\psi}\right)^N \ket{\Phi}.
\end{equation}

\subsection{Exact system--apparatus evolution}

The expression \eqref{eq:hdsv} for the system--apparatus evolution can be evaluated explicitly. The result was quoted in Ref.~\cite{Piacentini:2017:oo}, with some of the calculational steps provided in Ref.~\cite{Combes:2018:lm} for the general case of a state protection that does not necessarily project on the initial state. For completeness, here we will present the calculational steps for the case relevant to our analysis when the prepared and protected states are the same.

First, we rewrite the evolution operator $\op{U}$, Eq.~\eqref{eq:bvhb111}, as an infinite series, 
\begin{align}
  \op{U} &= \sum_{n=0}^\infty \left( -\frac{\I}{N}\right)^n \frac{(\op{\Pi}_+-\op{\Pi}_-)^n \otimes \op{P}^n}{n!}\notag  \\
         &= \op{\Pi}_+ \otimes \left( \sum_{n=0}^\infty \left( -\frac{\I}{N}\right)^n \frac{ \op{P}^n}{n!}\right) \notag  \\ & \quad + \op{\Pi}_- \otimes \left( \sum_{n=0}^\infty \left( \frac{\I}{N}\right)^n \frac{ \op{P}^n}{n!} \right) \notag \\
  &= \op{\Pi}_+ \otimes \exp\left[ -  \frac{\I}{N}\op{P} \right]+ \op{\Pi}_- \otimes \exp\left[ +  \frac{\I}{N}\op{P} \right],
\end{align}
where we have used that $(\op{\Pi}_+-\op{\Pi}_-)^n = \op{\Pi}_+^n + (-\op{\Pi}_-)^n$ and $\op{\Pi}_\pm^n=\op{\Pi}_\pm$. Then Eq.~\eqref{eq:hdsv} becomes \cite{Combes:2018:lm}
\begin{align}
 & \ket{\psi} \bra{Q}\left(\bra{\psi} \op{U} \ket{\psi}\right)^N \ket{\Phi} \notag \\ & \quad = \ket{\psi} \bra{Q} \left( r\, \E^{ -  \frac{\I}{N}\op{P} } + (1-r) \, \E^{  \frac{\I}{N}\op{P} }\right)^N \ket{\Phi} \notag \\
   &\quad = \ket{\psi} \sum_{n=0}^N {N \choose n} r^n (1-r)^{N-n} \bra{Q}\E^{ -  \frac{\I}{N}(2n-N)\op{P} } \ket{\Phi} \notag\\
&\quad \equiv \ket{\psi} f_{N,r}(Q),
\end{align}
where we have adopted the definition
\begin{equation}\label{eq:f}
f_{N,r}(Q) = \sum_{n=0}^N {N \choose n} r^n (1-r)^{N-n} \bra{Q}\E^{ -  \frac{\I}{N}(2n-N)\op{P} } \ket{\Phi}
\end{equation}
introduced in Ref.~\cite{Combes:2018:lm}. Note that $f_{N,r}(Q)$ represents the final, unnormalized pointer wave packet in the $Q$ representation.

Let the initial pointer state $\ket{\Phi} = \int_{-\infty}^\infty \Phi(Q) \ket{Q} \,\D Q$ be a Gaussian centered at zero,
\begin{equation}\label{eq:vjhsj}
  \Phi(Q) = \frac{1}{(2\pi \sigma^2)^{1/4}} \exp\left(-\frac{Q^2}{4\sigma^2}\right).
\end{equation}
The corresponding probability density $\mathcal{P}(Q)=\abs{\Phi(Q)}^2$ is a Gaussian of width $\sigma$, which is equal to the uncertainty in $Q$ given by $\sqrt{\langle Q^2 \rangle - \langle Q \rangle^2}$  \footnote{We note here that Ref.~\cite{Combes:2018:lm} used a slightly different definition of the Gaussian wave packet, for which the uncertainty differs from $\sigma$ by a factor of $\sqrt{2}$.}. Then Eq.~\eqref{eq:f} becomes \cite{Piacentini:2017:oo,Combes:2018:lm}
\begin{equation}\label{eq;bvhjsb}
f_{N,r}(Q) = \sum_{n=0}^N {N \choose n} r^n (1-r)^{N-n} \Phi[Q-(2n-N)/N].
\end{equation}

\subsection{Analyzing the final pointer state}

\begin{figure*}

\begin{flushleft}   \hspace*{2.9cm} (a) $\sigma=0.1$ \hspace*{4.65cm} (b) $\sigma=0.2$\end{flushleft}  

 \includegraphics[scale=0.5]{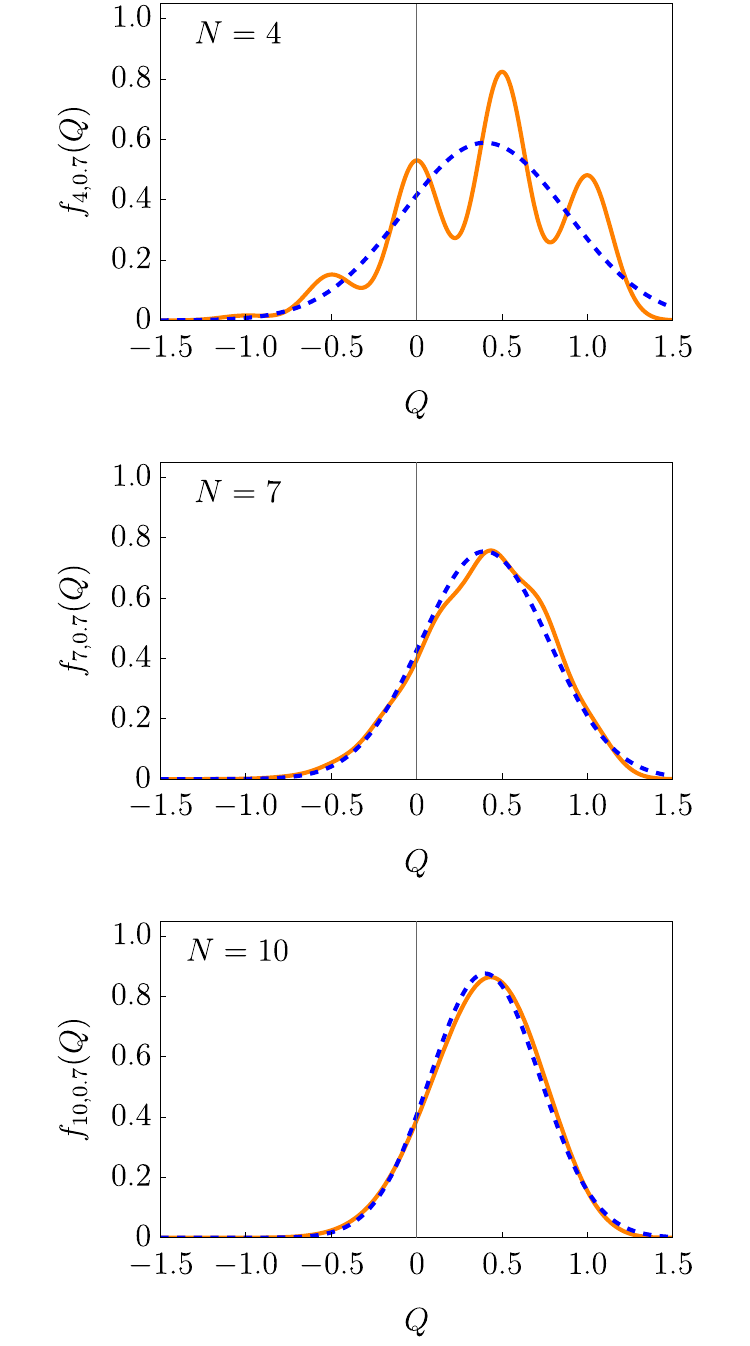}
 \includegraphics[scale=0.5]{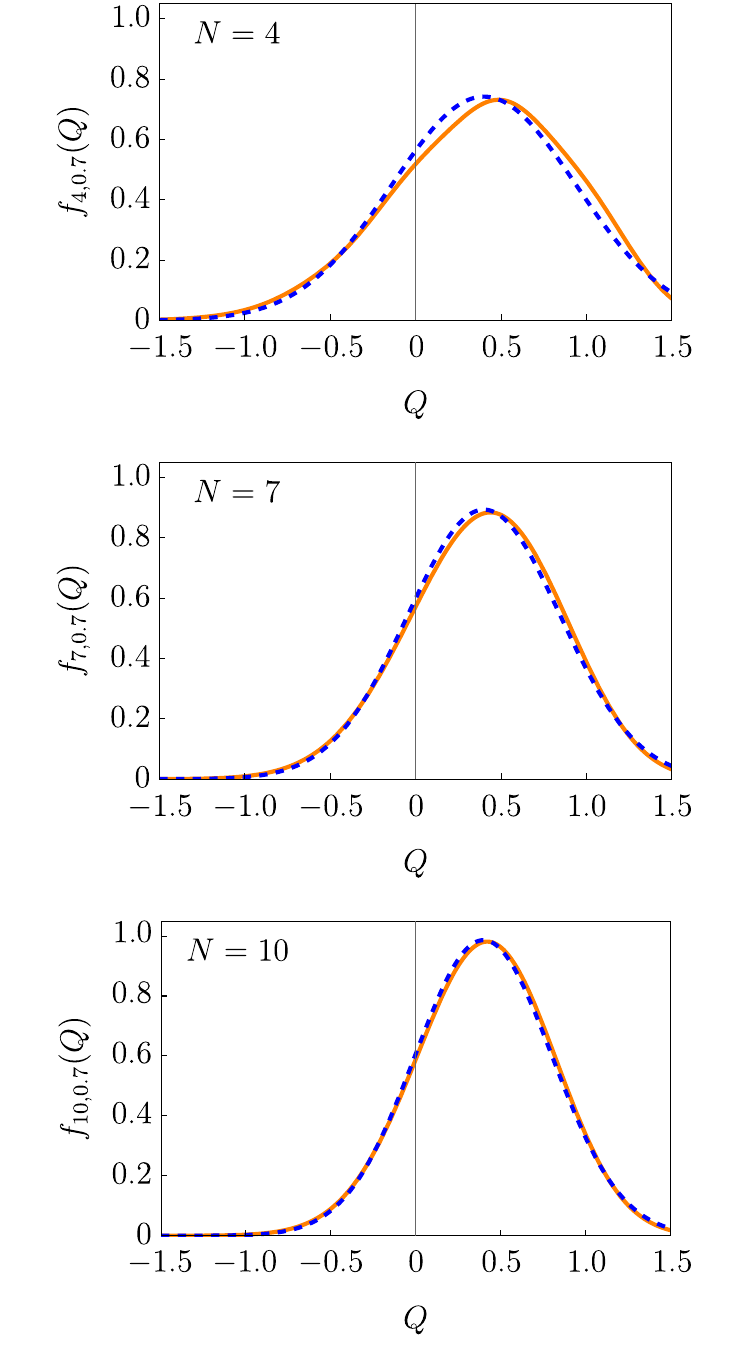}
 \caption{\label{eq:bincomp}Comparison of the exact expression, Eq.~\eqref{eq;bvhjsb}, for the final pointer wave packet (solid orange line) and the approximation, Eq.~\eqref{eq:vbsjhjshbvd}, based on the normal distribution (dashed blue line), for different values of $N$, with $r=0.7$ and (a) $\sigma=0.1$ and (b) $\sigma=0.2$. The plots show that whenever the measurement is reasonably weak [i.e., $(\sigma N)^{-1}\lesssim 1$], the normal distribution provides a good approximation.}
 \end{figure*}  

  \begin{figure}  
   \includegraphics[scale=.75]{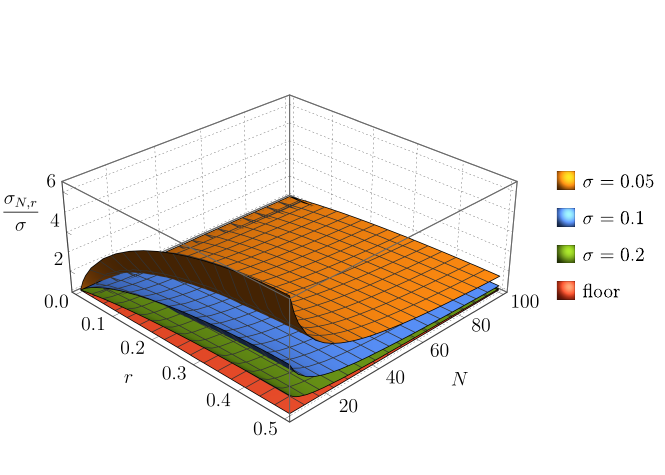}
   \caption{\label{fig:broaden}Broadening of the pointer wave packet during the protective measurement, as quantified in terms of the final width (uncertainty) $\sigma_{N,r}$ [see Eq.~\eqref{eq:jvshdsd}] relative to the initial width $\sigma$. The plot shows the ratio $\frac{\sigma_{N,r}}{\sigma}$ as a function of  $r$ (which specifies the state of the system) and $N$ (the number of Zeno stages, here between 5 and 100), for $\sigma=0.05$ (orange, top surface), $\sigma=0.1$ (blue, second from top), and $\sigma=0.2$ (green, third from top). The level plane (red, bottom surface) represents the floor $\frac{\sigma_{N,r}}{\sigma}=1$ corresponding to no broadening.}
 \end{figure}

To facilitate further evaluation of the final pointer wave packet, Eq.~\eqref{eq;bvhjsb}, we follow Ref.~\cite{Combes:2018:lm} and approximate the binomial distribution in Eq.~\eqref{eq;bvhjsb} using the normal distribution, 
\begin{multline}
{N \choose n} r^n (1-r)^{N-n} \\\approx \frac{1}{\sqrt{2\pi N r(1-r)}} \exp\left(- \frac{(n-Nr)^2}{2Nr(1-r)}\right),
\end{multline}
which improves as $N$ gets larger. This gives
\begin{align} \label{eq:vbsjhjshbvd}
f_{N,r}(Q) \approx \frac{\sqrt{\sigma}}{(2\pi)^{1/4} \sigma_{N,r}} \exp\left(- \frac{[Q-(2r-1)]^2}{4\sigma_{N,r}^2} \right),
\end{align}
where
 \begin{equation}\label{eq:jvshdsd}
\sigma_{N,r} = \sqrt{\frac{2r(1-r)}{N}+\sigma^2}
 \end{equation}
 is the width of the final probability density $\abs{f_{N,r}(Q)}^2$ for the pointer, and therefore represents the uncertainty of the pointer (in the $Q$ representation) at the end of the measurement \footnote{Equation~\eqref{eq:vbsjhjshbvd} differs from Eq.~(19) of Ref.~\cite{Combes:2018:lm} by a factor of 2 in the numerator due to the slightly differing definitions of the Gaussian wave packets.}. Figure~\ref{eq:bincomp} indicates that the normal distribution is an excellent approximation in the weak-measurement regime $N\sigma \gg 1$ relevant to a Zeno protective measurement. 

From Eq.~\eqref{eq:jvshdsd} we see that $\sigma_{N,r} \ge \sigma$, i.e., the pointer wave packet will always broaden (see Fig.~\ref{fig:broaden}) except in two limiting cases: (i) $N \rightarrow \infty$, i.e., for an infinitely weak protective measurement with infinitely many stages; (ii) $r=0$ or $r=1$, which correspond to the system being in the extremal states $\ket{-}$ and $\ket{+}$, respectively. The latter case can be understood by noting that if the system is in one of the extremal states, then the pointer shift will always be in the same direction, simply translating the pointer. For all other states, the pointer shift is a superposition of two opposite shifts (corresponding to the action of the unitary evolution operator on a superposition of  $\ket{+}$ and $\ket{-}$), which will distort (broaden) the wave packet. This broadening is most pronounced for an equal-weight superposition of $\ket{+}$ and $\ket{-}$, i.e., for $r=0.5$, in which case we have
\begin{equation}
\sigma_{N,r=0.5} = \sqrt{\frac{1}{2N}+\sigma^2}.
 \end{equation}

\begin{figure}  
\includegraphics[scale=.75]{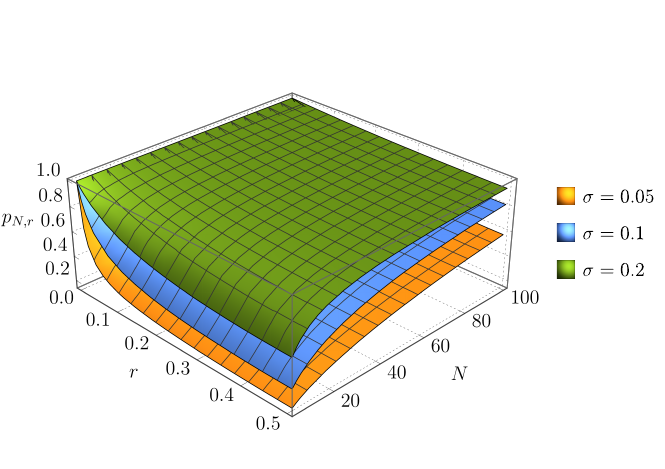}
\caption{\label{fig:survivalprob}Probability $p_{N,r}$, Eq.~\eqref{eq:vbhbdjh}, for the system to survive the $N$ protection stages, shown as a function of $N$ and $r$ (which specifies the state of the system), for $\sigma=0.05$ (orange, bottom surface), $\sigma=0.1$ (blue, middle surface), and $\sigma=0.2$ (green, top surface).}
\end{figure}

\begin{figure}  
\includegraphics[scale=.675]{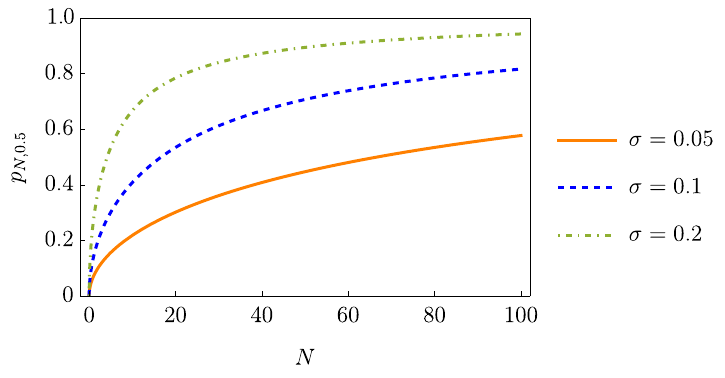}
\caption{\label{fig:survivalproblimit}Survival probability $p_{N,r}$, Eq.~\eqref{eq:vbhbdjh}, for a protective measurement of the state $\ket{\psi}=\frac{1}{\sqrt{2}}\left(\ket{+}+\ket{-}\right)$ (i.e., $r=0.5$), representing a lower bound on the probability for given $N$ and $\sigma$, shown for $\sigma=0.05$ (orange, solid line), $\sigma=0.1$ (blue, dashed), and $\sigma=0.2$ (green, dashed-dotted).}
\end{figure}

 The probability for the system to survive the $N$ protection stages is (see also Ref.~\cite{Combes:2018:lm})
 \begin{align}\label{eq:vbhbdjh}
 p_{N,r} &= \int_{-\infty}^\infty \abs{f_{N,r}(Q)}^2 \D Q = \frac{\sigma}{\sigma_{N,r}} \notag \\&= \left[ \frac{2r(1-r)}{N\sigma^2}+1\right]^{-1/2},
\end{align}
which is simply the inverse of the relative wave-packet broadening shown in Fig.~\ref{fig:broaden}, and is plotted in Fig.~\ref{fig:survivalprob}. For given $N$ and $\sigma$, the survival probability is lowest for the state with $r=0.5$, since for this state the broadening of the pointer wave packet is most pronounced. This bound is plotted in Fig.~\ref{fig:survivalproblimit}. It shows that, in the relevant weak-measurement regime $N\sigma \gg 1$, the survival probability is quite large (see also Ref.~\cite{Piacentini:2017:oo} for an analysis and discussion of survival probabilities in protective measurements). For example, for $\sigma=0.1$ and $N= 50$ Zeno stages (representing an only moderately weak measurement), the probability of surviving all Zeno stages, even in this worst-case scenario with $r=0.5$, is 0.71, which increases to 0.82 for $N= 100$ stages.

 \section{\label{sec:comp}Comparison of measurement uncertainties}

 We now use the above results to derive an expression that quantifies the performance of the protective measurement in terms of the ratio of the uncertainties for the protective measurement and the strong measurement. For concreteness, let us consider a photonic setting with $M$ initial photons, each prepared in the state $\ket{\psi}  = \cos\theta\ket{H}+\sin\theta \ket{V}$ (and thus $r=\cos^2\theta$), and a measurement of the linear polarization observable $\op{O} = \ketbra{H}{H} - \ketbra{V}{V}$. 

For a projective (strong) measurement with a beam splitter, we can assume that all $M$ photons are also detected. Therefore the uncertainty in the measured expectation value obtained from $M$ detected photons is the standard deviation of the mean, 
\begin{equation}\label{eq:kiki}
   u_\text{SM} = \frac{ \sqrt{ \langle \op{O^2} \rangle - \langle \op{O} \rangle^2}}{\sqrt{M}}= \frac{\sqrt{ 4r(1-r)}}{\sqrt{M}}= \frac{ \abs{\sin(2 \theta)}}{\sqrt{M}},
\end{equation}
where we have shown the result alternatively in terms of $r$ and $\theta$. 

 \begin{figure*}  
   \includegraphics[scale=.75]{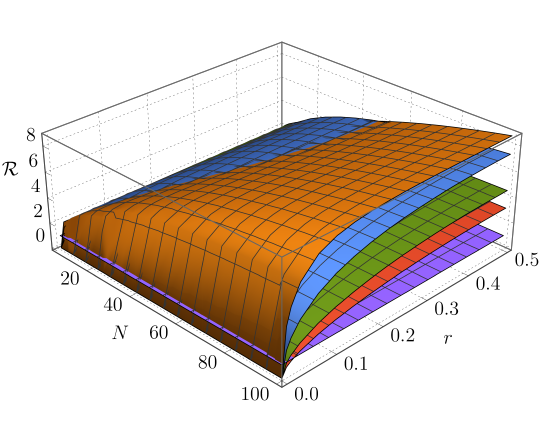}  \hspace*{.5cm} \includegraphics[scale=.75]{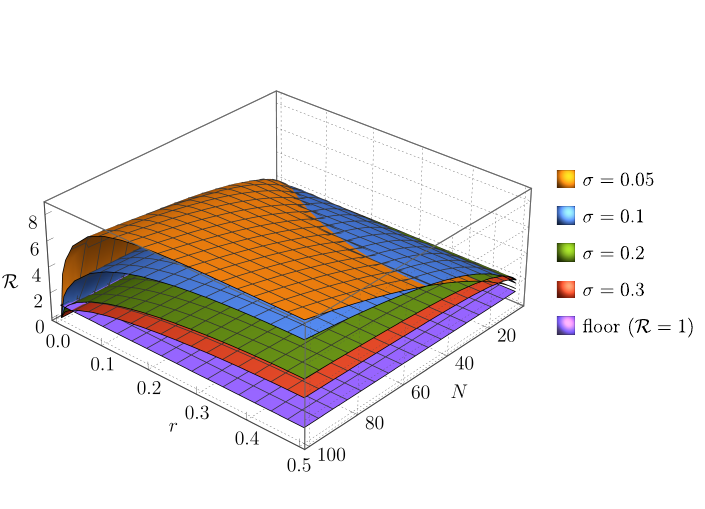} \hspace*{.5cm} \includegraphics[scale=.75]{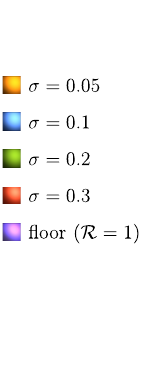}
   \caption{\label{fig:R}Performance of the protective measurement, shown in terms of the performance quantifier $\mathcal{R}$ given by Eq.~\eqref{eq:jvewshsviuft67hd}, as a function of the number $N$ of Zeno stages and the value $r$ that specifies the measured quantum state $\ket{\psi}  = \sqrt{r}\ket{+}+\sqrt{1-r} \ket{-}$. The two plots represent two different viewpoints of the same graph. The quantifier $\mathcal{R}$ represents the uncertainty in the protectively measured expectation value relative to the uncertainty in the expectation value obtained from a series of projective measurements.  Each surface in the graph corresponds to a different value of the initial width $\sigma$ of the pointer wave packet: $\sigma=0.05$ (orange, top surface), $\sigma=0.1$ (blue, second from top), $\sigma=0.2$ (green, third from top), and $\sigma=0.3$ (red, fourth from top).   The level plane (purple) represents the floor, $\mathcal{R}=1$, which corresponds to the same uncertainty (and thus the same performance) for both the protective and projective measurements. Since the plot is symmetric about $r=0.5$, we only show the region $0 \le r \le 0.5$.}
 \end{figure*}
 
For the protective measurement, only $p_{N,r} M$ photons are actually detected, where $p_{N,r}$ is the survival probability given by Eq.~\eqref{eq:vbhbdjh}. Thus the uncertainty is the standard deviation of the mean for this sample size,
\begin{equation}\label{eq:kiki2}
   u_\text{PM} = \frac{ \sigma_{N,r} }{\sqrt{p_{N,r} M}}.
\end{equation}
Following Refs.~\cite{Piacentini:2017:oo,Chen:2023:jn}, we quantify the relative measurement performance in terms of the ratio $\mathcal{R}=\frac{u_\text{SM}}{u_\text{PM}}$. Whenever $\mathcal{R}>1$, the protective measurement will be advantageous by virtue of the measured expectation value having smaller uncertainty. Using Eqs.~\eqref{eq:kiki} and \eqref{eq:kiki2} together with the expressions \eqref{eq:jvshdsd} and \eqref{eq:vbhbdjh} for $\sigma_{N,r}$ and $p_{N,r}$, we obtain our main result,
\begin{align}\label{eq:jvewshsviuft67hd}
  \mathcal{R} &= \frac{u_\text{SM}}{u_\text{PM}} = \frac{ \sqrt{p_{N,r}} \sqrt{ 4r(1-r)}}{\sigma_{N,r}} =
\frac{\sqrt{4 r(1-r)\sigma}}{\left[ \frac{2r(1-r)}{N}+\sigma^2\right]^{3/4}} \notag\\
  &= \frac{ \sqrt{\sigma}  \abs{\sin(2 \theta)}}{\left( \frac{\sin^2(2\theta)}{2N}+\sigma^2\right)^{3/4}},
\end{align}
which is plotted in Fig.~\ref{fig:R}. 

We see that the protective measurement is always advantageous ($R>1$) except when the state is very close to the extremal states $\ket{\pm}$ (i.e., if $r \ll 1$ or $r \approx 1$). Outside these extremal regions, the advantage persists even for a very small number $N$ of Zeno stages, although, as expected, the performance improves as $N$ is increased, i.e., as the pointer shift per stage is reduced. Another important  observation concerns the role played by the initial width $\sigma$ of the pointer wave packet. There is a subtle tradeoff between two competing effects. On the one hand, a smaller value of $\sigma$ makes the wave packet more sharply defined and thereby reduces the uncertainty in reading its center. On the other hand, a smaller value of $\sigma$ also makes the measurement stronger (since the overlap between wave packets shifted in adjacent Zeno stages will be reduced, making them more distinguishable), and hence the measurement leads to a greater state disturbance per Zeno stage. We see from Fig.~\ref{fig:R}, however, that the former influence clearly dominates the final uncertainty: it is generally advantageous to choose a smaller wave-packet width $\sigma$. This advantage diminishes as $N$ is reduced. This can be understood by noting that smaller $N$ means a larger average shift per Zeno stage, and therefore a narrower wave packet will be more affected (since the measurement has effectively become stronger).

Note that, in the limit of large $N$, we have $p_{N,r} \rightarrow 1$ and $\sigma_{N,r} \rightarrow \sigma$, and therefore Eq.~\eqref{eq:jvewshsviuft67hd} becomes
\[
  \mathcal{R} = \frac{ \sqrt{4r(1-r)}}{\sigma} = \frac{ \abs{\sin(2 \theta)} }{\sigma}.
  \]
As expected, in this case the uncertainty of the protective measurement is simply determined by the initial (and, in the limit of large $N$, unchanging) width of the pointer wave packet.

\section{\label{sec:discussion}Discussion}

The analytical results presented in this paper demonstrate that a measured expectation value will typically have a smaller uncertainty when it is obtained from a Zeno protective measurement than when it is obtained from a set of projective (strong) measurements on an ensemble. This finding expands on and confirms previous numerical \cite{Piacentini:2017:oo} and experimental \cite{Chen:2023:jn} results. The performance advantage can be amplified by making the pointer wave packet narrower (so that its center is more easily pinpointed) and by increasing the number $N$ of Zeno stages while proportionally reducing the coupling strength (in our model, this reduction is automatically built in by making the coupling proportional to $1/N$). 

Despite the smaller uncertainty that a Zeno protective measurement is able to achieve compared to projective measurements on an ensemble, one may wonder if this really makes protective measurements advantageous. After all, the protection procedure in a Zeno protective measurement is effectively a repreparation of the initial state and therefore it is reasonable to ask whether this may amount to having to know the initial state to begin with, in which case any expectation values could simply be calculated from this state. But all that is required for a Zeno protective measurement to be realized is that the protection stage projects on the same state as the initial preparation stage. For example, by employing an optical loop, a single polarizer could be used to both prepare and protect the photon state. Not only would the experimenter not need to know the setting of this polarizer, but he might not even be able to know it. For example, the polarizer setting could be chosen by a quantum random number generator, with the experiment sealed inside of a box, so that no information about the state that is prepared and protected is available to the experimenter. Another example is provided by the experimental realization of a photonic Zeno protective measurement described in Ref.~\cite{Chen:2023:jn}. There, the state protection is implemented by a polarization stabilizer, and it is impossible for the experimenter to know what the photon state is.

Of course, there are many situations for which the use of a Zeno protective measurement would be rather cumbersome for the task at hand, because it requires harnessing the state-preparation procedure to implement the protection, and it also requires multiple measurement and protection stages. Therefore, if the task is simply to measure expectation values on an ensemble, protective measurements are unlikely to challenge the primacy of strong measurements, such as those realized by sending photons through a beam splitter. But, like weak measurements in general, protective measurements offer unique features and possibilities that make them interesting in their own right and, as we have seen, can also provide a fundamental advantage in performance. 

Our model has neglected the influence of a potentially present environment on the system--apparatus evolution. Protective measurements of open quantum systems subject to decoherence-inducing interactions with an environment were previously studied by us in Ref.~\cite{Schlosshauer:2020:uu}. While that study used the model of adiabatic protective measurements rather than the Zeno protective measurements of our present paper, it is reasonable to expect that a similar study, using a similar Hamiltonian for the interaction with the environment, and using similar methods for solving the resulting system--apparatus--environment model, could also be applied to Zeno protective measurements. Given the mathematical similarities, one may anticipate that some of the main results of Ref.~\cite{Schlosshauer:2020:uu}---for example, the finding that the greatest impact of the environment is frequently not on the measured system, but on the behavior of the apparatus pointer---may also apply to Zeno protective measurements. However, given that in a Zeno protective measurement the system is repeatedly projected back onto its initial (pure) state, thereby disentangling it not only from the apparatus but also from any potentially present environment, one may conjecture that the influence of the environment on Zeno protective measurements may generally be less pronounced than for adiabatic protective measurements. Furthermore, in the experimentally relevant case of Zeno protective measurements implemented with photons \cite{Piacentini:2017:oo,Rebufello:2021:km,Chen:2023:jn}, environmental interactions are unlikely to play a significant role, since photons are largely immune to environmental decoherence in such experiments. Nonetheless, a rigorous study of Zeno protective measurements in the presence of a decoherence-inducing environment would constitute an interesting subject for future investigation.

We note that there is a close connection between Zeno protective measurements and a class of quantum walks in which the quantum particle is subject to a series of measurements \cite{Dhar:2015:hh,*Dhar:2015:kk,*Krovi:2006:aa,*Krovi:2007:ll,*Friedman:2017:uu,*Yin:2019:aa,*Liu:2020:aa,*Didi:2022:gg}. In such quantum walks, the unitary evolution of the quantum particle is interrupted at regular time intervals by repeated projective measurements performed on the particle, for example, in order to detect its position or to check whether it has reached a target site on the lattice. These measurements are quite analogous to the repeated state-protection steps in a Zeno protective measurement. 

Repeated state-projection steps have also been used to experimentally demonstrate the measurement of anomalous weak values on a single photon \cite{Rebufello:2021:jn}. This procedure may be considered a kind of generalization of the protective-measurement scheme. In both schemes, a series of weak measurements with intermediate protection steps is performed on a single system. In the Zeno protective-measurement scheme, the protection corresponds to a postselection onto the same state as the preselected state. In the weak-value measurement scheme of Ref.~\cite{Rebufello:2021:jn}, the postselected state is different, but it is subsequently rotated to match the preselected state, so that the state entering each measurement stage is identical to the initial state \footnote{This reduces to the standard protective-measurement scheme if the postselected state were to be chosen to be the same as the preselected state, in which case the weak value reduces to the expectation value.}. It would be interesting to investigate whether measuring weak values in this way may yield a performance advantage over the traditional way \cite{Aharonov:1988:mz,*Ritchie:1991:uu,*Hosten:2008:op,*Dixon:2009:ii,*Dressel:2014:uu} of measuring weak values from the statistics of single, weak measurements on an ensemble of identical pre- and postselected systems. 

\begin{acknowledgments}
We thank M.~Beck for useful comments. This work was funded by the National Science Foundation (Grant No.\ PHY-2109962). 
\end{acknowledgments}


%

\end{document}